# High-pressure synthesis of ultraincompressible hard rhenium nitride pernitride Re$_2$(N$_2$)N$_2$ stable at ambient conditions

Maxim Bykov[1,*], Stella Chariton[1], Hongzhan Fei[1], Timofey Fedotenko[2], Georgios Aprilis[2], Alena V. Ponomareva[3], Ferenc Tasnádi[4], Igor A. Abrikosov[4], Benoit Merle[5], Patrick Feldner[5], Sebastian Vogel[6], Wolfgang Schnick[6], Vitali B. Prakapenka[7], Eran Greenberg[7], Michael Hanfland[8], Anna Pakhomova[9], Hanns-Peter Liermann[9], Tomoo Katsura[1], Natalia Dubrovinskaia[2], Leonid Dubrovinsky[1]

[1]Bayerisches Geoinstitut, University of Bayreuth, Universitätstraβe 30, 95440 Bayreuth, Germany

[2]Material Physics and Technology at Extreme Conditions, Laboratory of Crystallography, University of Bayreuth, 95440 Bayreuth, Germany

[3]Materials Modeling and Development Laboratory, National University of Science and Technology 'MISIS', Moscow 119049, Russia

[4]Department of Physics, Chemistry and Biology (IFM), Linköping University, SE-58183 Linköping, Sweden

[5]Materials Science & Engineering, Institute I, Friedrich-Alexander-Universität Erlangen-Nürnberg (FAU) Martensstraβe. 5, D-91058 Erlangen, Germany

[6]Chair in Inorganic Solid State Chemistry, Department of Chemistry
University of Munich (LMU), Butenandtstraβe 5-13 (D), D-81377 Munich (Germany)

[7]Center for Advanced Radiation Sources, University of Chicago, 9700 South Cass Avenue, Argonne, IL 60437, USA

[8]European Synchrotron Radiation Facility, BP 220, 38043 Grenoble Cedex, France

[9]Photon Science, Deutsches Elektronen-Synchrotron, Notkestraβe 85, 22607 Hamburg, Germany.

*E-mail: maks.byk@gmail.com

**High-pressure synthesis can yield unique compounds with advanced properties, but usually they are either unrecoverable at ambient conditions or produced in quantity insufficient for properties characterization[1–4]. Here we report the synthesis of metallic, ultraincompressible (bulk modulus $K_0$ = 428(10) GPa), and very hard (nanoindentation hardness 36.7(8) GPa) rhenium (V) nitride pernitride $Re_2(N_2)N_2$. First it was obtained through a direct reaction between rhenium and nitrogen at 40 to 90 GPa in a laser-heated diamond anvil cell. The synthesis was scaled up through a reaction between rhenium and ammonium azide in a large-volume press at 33 GPa. Although metallic bonding is typically seen incompatible with intrinsic hardness, $Re_2(N_2)N_2$ turned to be at a threshold for superhard materials. Our work demonstrates a feasibility of surmounting conceptions common in material sciences.**

According to the approach formulated by Yeung *et al.*[4], the design of novel superhard materials should be based on the combination of a metal with high valence electron density with the first-row main-group elements, which form short covalent bonds to prevent dislocations. This conclusion was based on the synthesis of hard borides, such as $OsB_2$[5], $ReB_2$[6–8], $FeB_4$[9] or $WB_4$[10], whose crystal structures possess covalently bonded boron networks. Similar to boron, nitrogen as well can form covalent nitrogen-nitrogen bonds, but there are several factors, which make it difficult to synthesize nitrogen-rich nitrides. The large bond enthalpy of the triply bound $N_2$ molecule (941 kJ·mol$^{-1}$)[11] makes this element generally unreactive. In many reactions the activation barrier for $N_2$ bond breaking requires temperatures, which are higher than the decomposition temperatures of the target phases. $MN_x$ compounds with $x > 1$ are rarely available *via* direct nitridation reactions or ammonothermal syntheses[12,13]. Therefore, binary systems *M*-N systems are often limited to interstitial metal-rich nitrides. Usually, they are less compressible and have higher bulk moduli in comparison with pure metals due to the increasing repulsion between metal and nitrogen atoms, whereas their shear moduli are not always much different from those of metals.

Application of pressure is one way to increase the chemical potential of nitrogen and to stabilize nitrogen-rich phases.[14] Several transition metal dinitrides, $PtN_2$[15], $PdN_2$[16], $IrN_2$[17], $OsN_2$[17], $TiN_2$[18], $RhN_2$[19], $RuN_2$[20], $CoN_2$[21] and $FeN_2$[22], containing covalently bound dinitrogen units were recently synthesized in laser-heated diamond anvil cells (LHDACs) *via* reactions between elemental metal and nitrogen in a pressure range of 40-80 GPa. Although LHDAC is an efficient method to study high-pressure chemical reactions, it is challenging to scale up the synthesis. The search for suitable synthetic strategies, which would enable an appropriate reaction to be realized in a large volume press (LVP) instead of a LHDAC, is an important challenge for high-pressure chemistry and materials sciences. In this study, focusing on the high-pressure synthesis of novel nitrogen-rich phases in the Re-N system and the development of new synthetic strategies, we resolved this problem for a previously unknown rhenium nitride $ReN_2$ with unusual crystal chemistry and unique properties.

Direct reactions between rhenium and nitrogen were studied by Friedrich *et al.*[23], who synthesized two interstitial rhenium nitrides $Re_3N$ at 13 GPa and 1700 K, and $Re_2N$ at 20 GPa and 2000 K. Both compounds have exceptionally large bulk moduli exceeding 400 GPa (as measured upon compression in a non-hydrostatic medium[23]), but only moderate shear moduli as expected for interstitial compounds[24]. Kawamura *et al.*[25] reported synthesis of $ReN_2$ with $MoS_2$ structure type ($m$-$ReN_2$) in a metathesis reaction between $Li_3N$ and $ReCl_5$ at 7.7 GPa. Subsequently, Wang *et al.*[26] suggested, based on the first-principle calculations, that $m$-$ReN_2$ is unstable and 'real stoichiometric' $ReN_2$ should have monoclinic $C2/m$ symmetry and transform to the tetragonal $P4/mbm$ phase above 130 GPa. However, this suggestion has not been proven experimentally as yet. Recently Bykov *et al.* reported a novel inclusion polynitrogen compound $ReN_8 \cdot xN_2$ synthesized from elements at 106 GPa[27], but the region of ~35-100 GPa still remains completely unexplored for the Re-N system.

To fill this gap in pressure and temperature space, we have studied chemical reactions between Re and nitrogen and other reagents, such as sodium azide $NaN_3$ and ammonium azide $NH_4N_3$, in LHDACs in a range of 26 – 87 GPa at temperatures of 2000 – 2500 K (Table 1, Experiments #1 through #6). The reactions products typically contained numerous single-crystalline grains of several rhenium nitride phases (Table 1), which were identified using synchrotron single-crystal X-ray diffraction (Supplementary Figure 1, Supplementary Table 1).

**Table 1. Summary of syntheses**

| Experiment, | Technique | Reagents | Pressure (GPa) | Temperature (K) | Products |
|---|---|---|---|---|---|
| 1 | LHDAC | $Re + N_2$ | 42 | 2200(300) | $ReN_2 + Re_2N + ReN_{0.6}$ |
| 2 | LHDAC | $Re + N_2$ | 49 | 2200(300) | $ReN_2 + Re_2N$ |
| 3 | LHDAC | $Re + N_2$ | 71 | 2500(300) | $ReN_2 + Re_2N$ |
| 4 | LHDAC | $Re + N_2$ | 86 | 2400(300) | $ReN_2 + Re_2N$ |
| 5 | LHDAC | $Re + NaN_3$ | 29 | 2000(300) | $NaReN_2 + Re_2N$ |
| 6 | LHDAC | $Re + NH_4N_3$ | 43 | 2200(300) | $ReN_2 + ReN_{0.6} + Re_2N$ |
| 7 | LVP | $Re + NH_4N_3$ | 33 | 2273(100) | $ReN_2 + Re_2N$ |

A direct reaction between Re and $N_2$ (Table 1) resulted in the synthesis of three rhenium nitrides $ReN_2$, $Re_2N$, and $ReN_{0.6}$, two of which ($ReN_2$ and $ReN_{0.6}$) have never been observed before. The third phase identified in these experiments, $Re_2N$ ($P6_3/mmc$), has previously been reported[23]. After a stepwise decompression of the sample obtained in Experiment #1 down to the ambient pressure, all of the three phases (ReN$_2$, ReN$_{0.6}$, Re$_2$N) were found to remain intact even after one month (Supplementary Figure 4). Crystal structure analysis of ReN$_{0.6}$ showed that it has a defect WC structure type (space group $P\bar{6}m2$) (for details on ReN$_{0.6}$ see Supplementary Information; Supplementary Figures 2, 3, Supplementary Table 2).

Analysis of the crystal structure of ReN$_2$ revealed its unusual crystal-chemistry. Figure 1 shows the crystal structure of ReN$_2$, which is built of distorted ReN$_7$ capped trigonal prisms (Figure 1d) and contains both N-N units (dumbbells) (Figure 1f) and discrete N atoms (N2) (Figure 1e) in an atomic ratio 1:1. The N1-N1 dumbbells are located in a trigonal antiprism formed by Re atoms (Figure 1f), while discrete N2 atoms have a tetrahedral coordination by Re (Figure 1e). The N1-N1 bond length ($d_{N1-N1}$ = 1.412(16) Å at ambient conditions) suggests that the N$_2$ unit should be considered as a pernitride anion N$_2^{4-}$. Therefore, ReN$_2$ is a rhenium nitride pernitride and its crystal-chemical formula is Re$^{+V}_2$[N$^{-II}_2$][N$^{-III}$]$_2$.

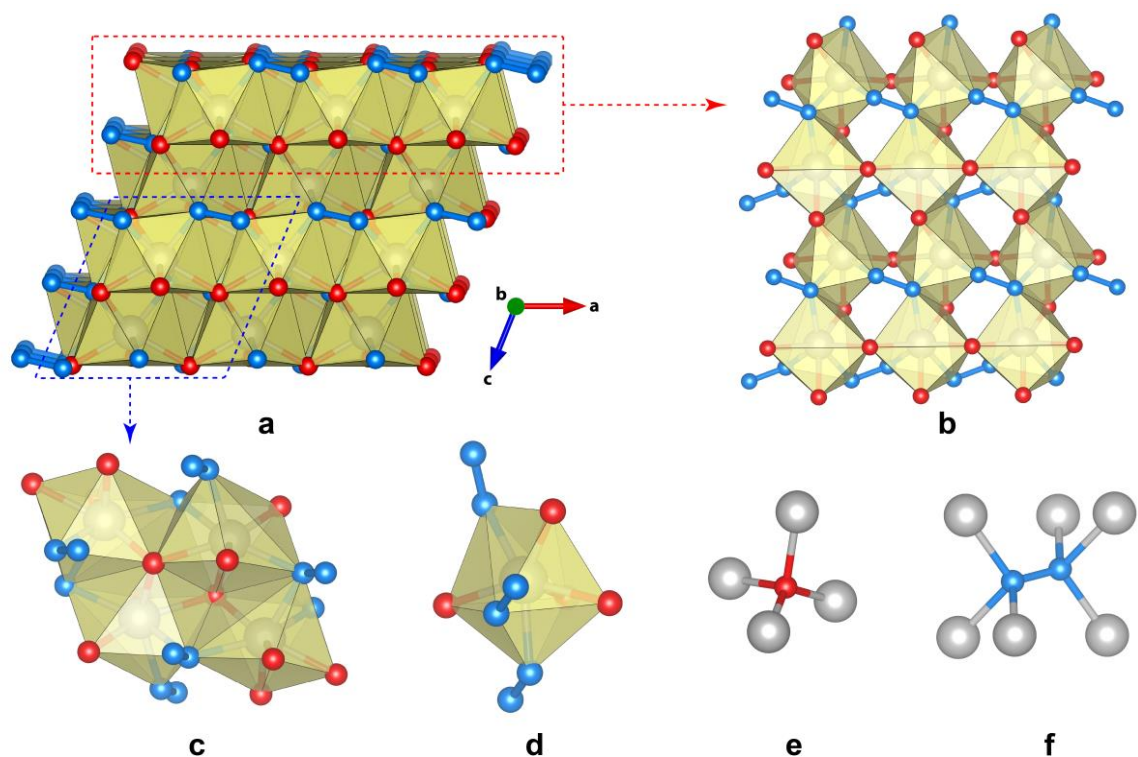

**Figure 1. Fragments of the crystal structure of Re$_2$(N$_2$)(N)$_2$ at ambient conditions. Re atoms – gray, N1 atoms – blue, N2 atoms – red.** ReN$_2$ crystallizes in the space group $P2_1/c$ (No. 14) with $a$ = 3.6254(17), $b$ = 6.407(7), $c$ = 4.948(3) Å, $\beta$ = 111.48(6)°. Rhenium and nitrogen atoms occupy crystallographic sites 4$e$: Re [0.35490(11), 0.34041(8), 0.19965(8)], N1 [0.194(2), 0.038(2), 0.311(19)], N2 [0.259(3), 0.6381(18), 0.024(2)]. Full crystallographic information is given in the supplementary crystallographic information file and in the Supplementary Tables 3 and 4. (a) The projection of the crystal structure along the *b*-axis. (b,c) Fragments of the crystal structure of ReN$_2$ showing how ReN$_7$ polyhedra are connected with each other. (d) Separate ReN$_7$ coordination polyhedron. (e) Coordination of N2 atoms. (f) Coordination of N1-N1 dumbbells.

The compressibility of ReN$_2$ was measured on the sample #2 (Table 1), which was synthesized at 49 GPa, then decompressed down to ambient conditions, and re-loaded into another DAC with a neon pressure-transmitting medium, which provides much better hydrostaticity of the sample environment than nitrogen[28]. The sample was first characterized using single-crystal XRD at ambient conditions. On compression, the lattice parameters were extracted from the powder XRD data (Figure 2*a*,*b*; Supplementary Figures 4-6; Supplementary Table 5). The pressure-volume dependence was described using the third-order Birch-Murnaghan equation of state[29] with the following fit parameters: $V_0$ = 107.21(4) Å$^3$, $K_0$ = 428(10) GPa, $K'$ = 1.6(5). Figure 2*c* shows a plot of correlated values of $K_0$ and $K'$ to different confidence levels. The bulk modulus $K_0$ lies within the range of 410-447 GPa at the 99.73% confidence level. Thus, $K_0$ of ReN$_2$ is larger than that of any compound in the Re-N system and is comparable to that of diamond ($K_0$ = 440 GPa) and IrN$_2$ ($K_0$ = 428(12) GPa)[17]. Among very incompressible pernitrides of transition metals, ReN$_2$ is the only compound, in which the metal atom has oxidation state (+V) higher than (+IV). The increased ionicity of Re-N chemical bonding, compared to those in other pernitrides may, therefore, play a role in the enhancement of the bulk modulus of ReN$_2$ in comparison to OsN$_2$, PtN$_2$ and TiN$_2$. This is in agreement with the trend recently proposed by Bykov *et al.*[22], who showed that the bulk moduli of dinitrides MN$_2$ increase with the increase of the oxidation state of metal atoms from +II to +IV.

More detail characterization of physical properties of ReN$_2$, such as hardness, electrical conductivity *etc.* require a sample to be at least a few tens of microns in size that is much larger than can be synthesized in a LHDAC. The large volume press technique enables the synthesis of such a sample, but precludes from using N$_2$ as a reagent. First, the amount of nitrogen, which can be sealed in a capsule along with Re, is insufficient for the desired reaction yield; second, unavoidable deformation of the capsule upon compression may potentially lead to the loss of nitrogen. Therefore, a solid source of nitrogen had to be found and we tested sodium and ammonium azides, NaN$_3$ and NH$_4$N$_3$, as potential precursors in LHDACs (Experiments #5, #6, Table 1) (for a discussion regarding the choice of the solid reagents see Supplementary Information). The experiment with NaN$_3$ (Experiment #5) did not result in the synthesis of ReN$_2$. The major product of the reaction was NaReN$_2$ (Supplementary Figure 7), whose lattice parameters turned out to be very close to those reported for *m*-ReN$_2$ by Kawamura *et al.*[25], that might suggest that the material described in Ref. 25 as rhenium nitride indeed could be a different compound (for a related discussion see Supplementary Information). The experiment in LHDAC with NH$_4$N$_3$ as a source of nitrogen resulted in the synthesis of ReN$_2$ among other products (Experiment #6, Table 1).

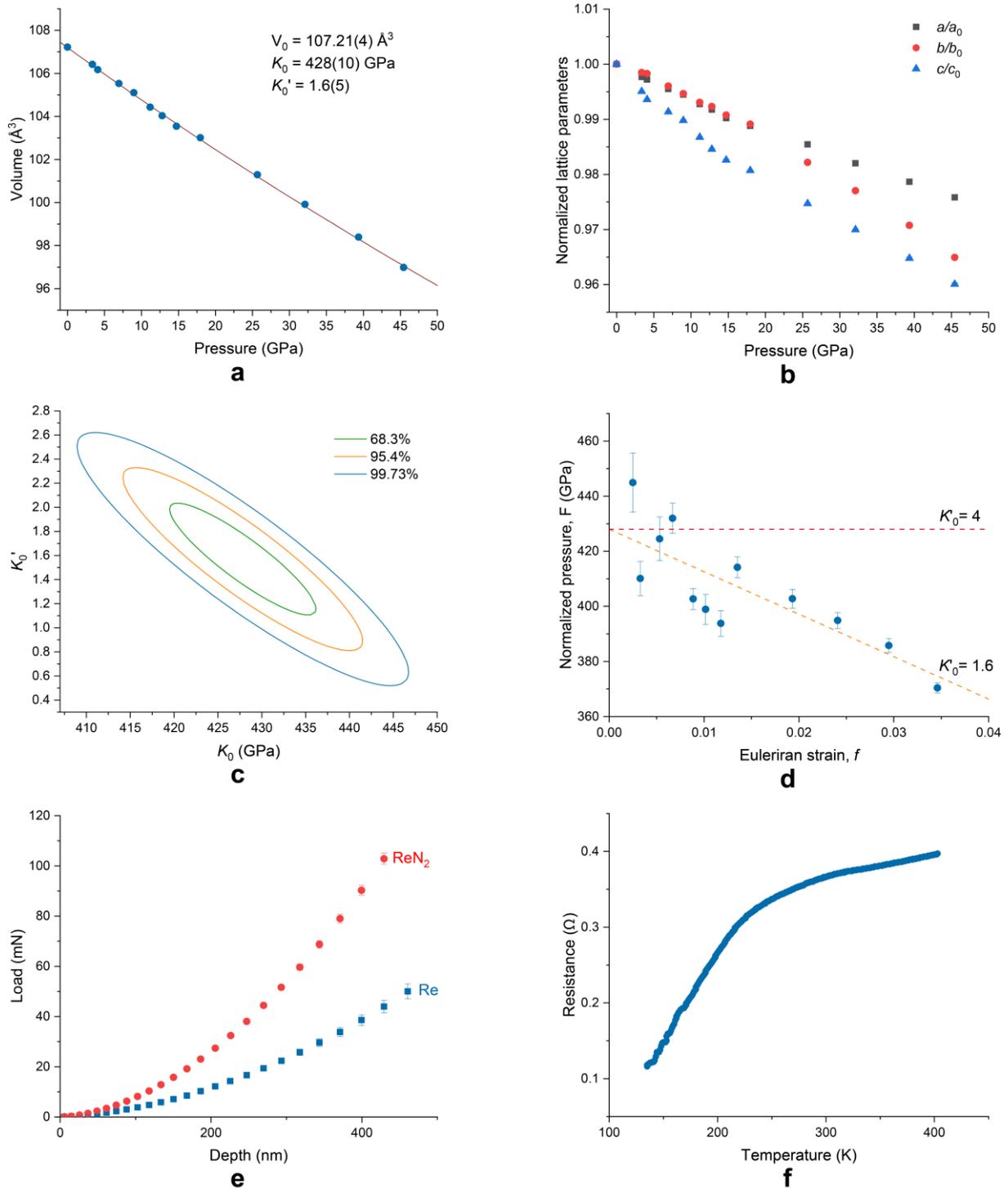

**Figure 2**. **Physical propeties of $Re_2(N_2)N_2$** (a) Pressure-dependence of the unit-cell volume and (b) normalized lattice parameters of $ReN_2$. (c) Plot of the correlated values of $K_0$ and $K_0'$ to different confidence levels of 68.3%, 95.4% and 99.73%, respectively. (d) An F-f plot based on Birch-Murnaghan EoS. (e) Averaged indentation load-displacement data. The error bars correspond to the standard deviation between 16 single measurements at different locations. (f) Temperature dependence of the electrical resistance of the $ReN_2$ sample at ambient pressure.

Based on results of this experiment in DAC, we explored a possibility to scale up the synthesis of ReN$_2$ in a multianvil LVP at 33 GPa and 2273 K via a reaction between rhenium and ammonium azide (Experiment #7, Supplementary Figure 8). The product of the reaction was a mixture of Re$_2$N and ReN$_2$. Each phase was separated (Supplementary Figure 9) and characterized using single-crystal X-ray diffraction. A phase-pure polycrystalline sample of ReN$_2$ (70 × 60 × 50 μm$^3$), which was synthesized in the LVP, was used for nanoindentation hardness and electrical resistance measurements. Nanoindentation was performed using a nanoindenter equipped with Berkovich diamond tip and featuring continuous stiffness measurement capabilities. The average hardness and Young's modulus measured between 200 and 400 nm depths are 36.7(8) GPa and 493(14) GPa respectively (Figure 2e, Table 3). The hardness approaching 40 GPa, a threshold for superhard materials, and extreme stiffness comparable with diamond makes mechanical properties of ReN$_2$ exceptional in the row of metal nitrides. Due to the directional N-N bonding, the hardness of ReN$_2$ is higher than that of known interstitial transition metal nitrides (δ-NbN – 20 GPa, HfN – 19.5 GPa, ZrN – 17.4 GPa[30], CrN -17 GPa [4] *etc.*). Most transition metal pernitrides MN$_2$ that are metastable at ambient conditions are expected to be very hard compounds too, however they were never obtained in a quantity sufficient for the hardness measurements[31,32].

The electrical resistance of ReN$_2$ as a function of temperature was measured at ambient pressure on a sample with the dimensions of about 70×60×50 μm$^3$. The results of the measurements in the range of 150 K to 400 K are shown in Figure 2f. Electrical resistivity of metals increases with temperature and this is the case for ReN$_2$. The shape of the resistance – temperature curve (Figure 2f) is reproducible as confirmed in a number of independent measurements on the same sample with re-glued electrical contacts.

To confirm the experimentally observed peculiarities of ReN$_2$ and to gain deeper insights into the mechanical and electronic properties of this compound, we performed theoretical calculations based on the density functional theory. First, we considered the crystal structure of ReN$_2$. We carried out the full structure optimization for the compound at ambient pressure and found that calculations and experiment are in very close agreement (Supplementary Table 3). Calculated elastic constants for ReN$_2$ (Table 2) fulfill the mechanical stability conditions[33], and calculated phonon dispersion relations (Figure 3a) show only real frequencies confirming its dynamic stability. Theoretically calculated N1−N1 vibrational frequency form a localized band giving rise to a peak of the phonon density of states at ~1031 cm$^{-1}$. This vibrational behavior is similar to other pernitrides[15,18,34]. The metallic nature of the material confirmed by our calculations of the electronic density of states (DOS) (Figure 3c).

Calculated vibrational and electronic properties of N1-N1 unit confirm that it is a pernitride anion $N_2^{4-}$. On the contrary, electronic and vibrational properties of N2 atoms (Figure 3a,c) are quite distinct from those of N1, providing strong support to the experimental observation of the crystal chemistry of $ReN_2$, which is unique for transition metals pernitrides.

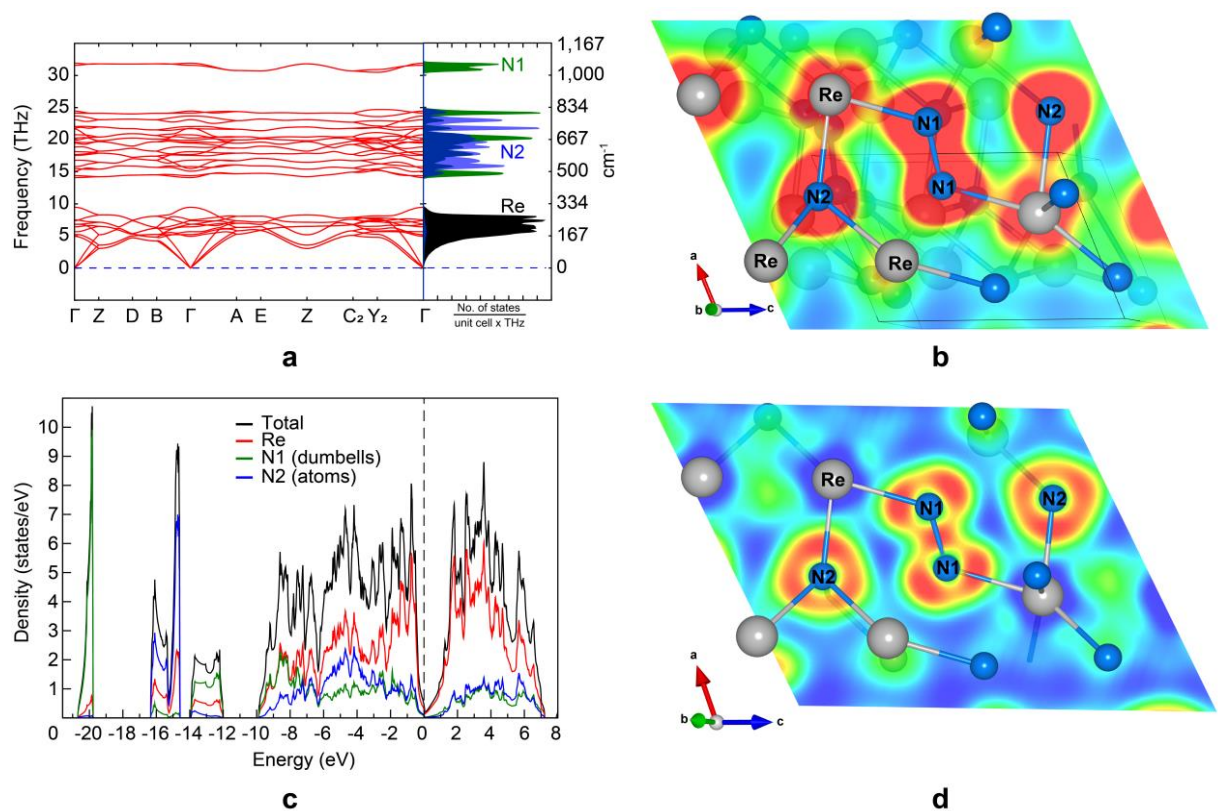

**Figure 3**. **Phonon and electronic structure calculations for $ReN_2$.** Calculated phonon dispersion relations (a), charge density map (b), densities of states (c) and electron localization funciton (d) for $ReN_2$ at ambient conditions.

The unique chemistry of the compound is essential for understanding its superior mechanical properties. The bulk modulus, calculated theoretically using Voigt-Reuss-Hill approximation (413.5 GPa)[35] is in a good agreement with the experiment (428(10) GPa), confirming that $ReN_2$ can be characterized as highly incompressible material. At the same time, the value of the Poisson coefficient is close to 0.25, and relatively high ratio between share and bulk elastic moduli indicates substantial degree of covalence in $ReN_2$ chemical bonding. A direct calculation of the charge density map (Figure 3b) confirms the expectation. One sees a formation of covalent bonds between two N1 atoms. It is of single bond character with very high degree of electron localization (Figure 3d) typical for a pernitride anion $N_2^{4-}$ in other transition metal pernitrides,[36] and incompressible $N_2^{4-}$ is supposed to contribute to very low compressibility of the materials. The covalent bond between Re and N2 atoms is formed by substantially less localized electrons (compare Fig. 3b and Fig 3d). Indeed, measured temperature dependence of the electrical resistance (Figure 2f) and the estimated resistivity (~4·10$^{-6}$ Ω·m - 16·10$^{-6}$ Ω·m in a temperature range 150-400 K) are in agreement with the theoretical conclusion and the description of ReN$_2$ as a metal. The formation of the covalent bonds between Re and N2 atoms indicates strong hybridization between the electronic states of the atoms. The calculated electronic DOS (Figure 3c) shows the presence of the pseudogap between occupied, predominantly bonding states of Re and unoccupied non-bonding and anti-bonding states. According to S.-H. Jhi *et al.*[37], such features optimize electronic contribution to hardness enhancement in transition-metal carbonitrides, which can also explain very high hardness of ReN$_2$. Thus, the formation of strong covalent bond between Re and N2 atoms, a unique feature of the material synthesized in this work in comparison with known transition metal pernitrides, appears to be essential for its spectacular mechanical and electronic properties.

**Table 2. Calculated elastic constants $C_{ij}$ (GPa), bulk modulus $B$ (GPa), shear modulus $G$ (GPa), Young's modulus $E$ (GPa) and Poisson's ratio ($v$) of ReN$_2$**

| $C_{11}$ | $C_{12}$ | $C_{13}$ | $C_{15}$ | $C_{22}$ | $C_{23}$ | $C_{25}$ | $C_{33}$ | $C_{35}$ |
|---|---|---|---|---|---|---|---|---|
| 869.51 | 230.73 | 261.47 | 51.01 | 748.93 | 251.83 | 26.67 | 648.06 | 16.61 |
| $C_{44}$ | $C_{46}$ | $C_{55}$ | $C_{66}$ | $B$ | $G$ | $E$ | $v$ | |
| 257.43 | 35.91 | 299.94 | 266.34 | 413.5 | 262 | 650 | 0.24 | |

**Table 3. Mean hardness and Young's modulus measured by nanoindentation in the 200-400 nm depth range. The error estimate corresponds to the standard deviation between 16 different locations.**

| Material | Hardness (GPa) | Young's modulus (GPa) |
|---|---|---|
| ReN$_2$ | 36.7(8) | 493(14) |
| Re | 10.9(6) | 424(12) |

To summarize, in the present work we have synthesized a novel transition metal nitride ReN$_2$ (Re$^{+V}_2$(N$^{-II}_2$)(N$^{-III}$)$_2$) with the unique crystal structure and outstanding properties. The structure with Re atoms in the high oxidation state +V features both discrete nitride and pernitride ions. A combination of the high electron density of the transition metal with interstitial nitride anions and covalently bound pernitride units makes this compound ultraincompressible and extremely hard at the same time. The developed method for scaling up the synthesis of ReN$_2$ in a LVP using ammonium azide as a nitrogen precursor may be applied for producing other transition metal nitrides. We demonstrated the complete route for materials development from screening experiments in diamond anvil cells to the synthesis of samples large enough for physical property measurements. It is not only our results *per se* that are important, but the fact that the development and synthesis of the new material was realized contrary to established concepts and should encourage further theoretical and experimental works in the field.

**Methods**

*Synthesis of Re-N phases in laser-heated diamond anvil cells*

In all synthesis experiments a rhenium powder (Sigma Aldrich, 99.995 %) was loaded into the sample chamber of a BX90 diamond anvil cell (Boehler-Almax anvils, 250-μm size). In four experiments the chamber was filled with nitrogen at 1.5 kbar that served as a pressure-transmitting medium and as a reagent. In two experiments, the chamber was filled either with ammonium azide NH$_4$N$_3$ or with sodium azide NaN$_3$. Pressure was determined using the equation of state of rhenium[38–40]. The compressed sample was heated using the double-sided laser-heating system installed at the Bayerisches Geoinstitut (BGI), University of Bayreuth, Germany. Successful syntheses were performed at 40, 42, 49, 71 and 86 GPa at temperatures of 2200-2500 K (Table 1).

*Synthesis of Re-N phases in the large-volume press*

High-pressure synthesis was performed using a Kawai-type multi-anvil apparatus IRIS15, installed at the BGI[41]. The NH$_4$N$_3$ sample (0.5 mm thickness, 0.8 mm in diameter) was sandwiched between two layers of Re powder (0.1 mm thick, 0.8 mm in diameter) and between two tubes of dense alumina in a Re capsule, which also acted as a heater. The capsule was placed in a 5 wt% Cr$_2$O$_3$-doped MgO octahedron with a 5.7 mm edge that was used as the pressure medium. The assembly scheme is given in the Supplementary Figure 11. Eight tungsten carbide cubes with 1.5 mm truncation edge lengths were used to generate high pressures. The assembly was pressurized at ambient temperature to 33 GPa, following the calibration given by Ishii *et al.*[41] and then heated to ~2273(100) K within 5 minutes and immediately quenched after the target temperature was reached. The assembly was then decompressed during 16 hours.

*Synthesis of NH$_4$N$_3$*

Ammonium azide, NH$_4$N$_3$ was obtained by the metathesis reaction between NH$_4$NO$_3$ (2.666 g, 33mmol, Sigma-Aldrich, 99.0%) and NaN$_3$ (2.165 g, 33 mmol, Acros Organics, Geel,

Belgium, 99%) in a Schlenk tube. By heating from room temperature to 170ºC in a glass oven and annealing for 7.5 h at 170ºC and then for 12h at 185ºC, $NH_4N_3$ precipitated at the cold end of the tube separated from $NaNO_3$, which remained at the hot end during the reaction[42].

*Compressibility measurements*

For the compressibility measurements the sample synthesized at 49 GPa and 2200 K (Experiment #2) was quenched down to ambient pressure and re-loaded into another diamond anvil cell. The sample chamber was then filled with Ne that served as a pressure-transmitting medium. A powder of gold (Sigma Aldrich, 99.99 %) was placed into the sample chamber along with the sample and used as a pressure standard[43]. The sample was then compressed up to ~45 GPa in 13 steps. At each pressure point we have collected powder X-ray diffraction data.

*Synchrotron X-ray diffraction*

High-pressure single-crystal and powder synchrotron X-ray diffraction studies of the reaction products were performed at the beamlines P02.2 (DESY, Hamburg, Germany)[44], ID15B (ESRF, Grenoble, France) and GSECARS beamline (APS, Argonne, USA). The following beamline setups were used. P02.2: $\lambda = 0.29$ Å, beam size ~2×2 μm$^2$, Perkin Elmer XRD 1621 detector; ID15B: $\lambda = 0.41$, beam size ~10×10 μm$^2$, Mar555 flat panel detector; GSECARS: $\lambda = 0.2952$ Å, beam size ~3×3 μm$^2$, Pilatus CdTe 1M detector. For the single-crystal XRD measurements samples were rotated around a vertical ω-axis in a range ±38°. The diffraction images were collected with an angular step $\Delta\omega = 0.5°$ and an exposure time of 1s/frame. For analysis of the single-crystal diffraction data (indexing, data integration, frame scaling and absorption correction) we used the *CrysAlis$^{Pro}$* software package. To calibrate an instrumental model in the *CrysAlis$^{Pro}$* software, *i.e.*, the sample-to-detector distance, detector's origin, offsets of goniometer angles, and rotation of both X-ray beam and the detector around the instrument axis, we used a single crystal of orthoenstatite (($Mg_{1.93}Fe_{0.06}$)($Si_{1.93}$, $Al_{0.06}$)$O_6$, *Pbca* space group, $a = 8.8117(2)$, $b = 5.18320(10)$, and $c = 18.2391(3)$ Å). The same calibration crystal was used at all the beamlines.

Powder diffraction measurements were performed either without sample rotation (still images) or upon continuous rotation in the range ±20°ω. The images were integrated to powder patterns with Dioptas software[45]. Le-Bail fits of the diffraction patterns were performed with the TOPAS6 software.

*In-house X-ray diffraction*

Ambient-pressure single-crystal XRD datasets were collected with a high-brilliance Rigaku diffractometer (Ag *Kα* radiation) equipped with Osmic focusing X-ray optics and Bruker Apex CCD detector in the BGI.

*Structure solution and refinement*

The structure was solved with the ShelXT structure solution program[46] using intrinsic phasing and refined with the Jana 2006 program[47]. CSD-1897795 contains the supplementary crystallographic data for this paper. These data can be obtained free of charge from FIZ Karlsruhe *via* www.ccdc.cam.ac.uk/structures

*Nanoindentation*

Nanoindentation was performed using a Nanoindenter G200 platform (KLA-Tencor, Milpitas, CA, USA), equipped with a Berkovich diamond tip (Synton MDP, Nidau, Switzerland) and featuring the continuous stiffness based method (CSM)[48]. Each sample was indented at 16 different locations separated by a distance of at least 10 µm, so that their plastic zones did not overlap. For each measurement, loading was performed at a constant strain-rate of 0.025 s$^{-1}$ up to a maximal indentation depth of at least 400 nm. A 2 nm large oscillation superimposed on the loading signal allowed continuously measuring the contact stiffness. The acquired data were evaluated using the Oliver-Pharr method [49,50]. To this purpose, the diamond punch geometry was calibrated from 1000 nm deep references measurements in fused silica and the machine frame stiffness value was refined so as to obtain a constant ratio between stiffness squared and load during indentation of the samples. The conversion of the reduced moduli to a uniaxial Young's moduli was performed assuming a Poisson's ratios of 0.24 and 0.29, respectively for ReN$_2$ and Re [51].

*Temperature-dependent resistance measurements*

The resistance of the sample was measured by four-probe method passing a constant DC 90mA current through the sample and measuring both current and voltage drop across the sample. Temperature was measured using the S-type thermocouple.

*Theoretical calculations*

The *ab initio* electronic structure calculations of ReN$_2$ (12 atoms), ReN (2 atoms) and ReN$_x$ (2×3×2 supercell) were performed using the all electron projector-augmented-wave (PAW) method[52] as implemented in the VASP code[53–55]. Among the tested exchange-correlation potentials (PBE[56], PBEsol[57], AM05[58]) the PBEsol approximation has resulted into the best agreement between the derived experimental and theoretical equation of state. Convergence has been obtained with 700 eV energy cutoff for the plane wave basis and a (18×10×14) Monkhorst-Pack k-points [59] type sampling of the Brillouin zone for ReN$_2$. Gaussian smearing technique was chosen with smearing of 0.05 eV. The convergence criterion for the electronic subsystem has been chosen to be equal to 10$^{-4}$ eV for two subsequent iterations, and the ionic relaxation loop within the conjugated gradient method was stopped when forces became of the order of $10^{-3}$ eV/Å. The elastic tensor $C_{ij}$ has been calculated from the total energy applying (+/-) 1 and 2% lattice distortions. The Born mechanical stability conditions have been verified using the elastic constants. The phonon calculations have been performed within quasiharmonic approximation at temperature T = 0 K using the finite displacement approach implemented into PHONOPY software[60]. Converged phonon dispersion relations have been achieved using a (3×3×3) supercell with 324 atoms and (5×5×5) Monkhorst-Pack *k*-point sampling.

**Author contributions**

M.B., L.D., and ND designed the research, M.B., L.D., N.D., I.A.A. wrote the manuscript, M.B, L.D., S.C, T.F, G.A., V.B.P., E.G., M.H., A.P. H.-P.L. performed X-ray diffraction experiments, M.B. analyzed the X-ray diffraction data, F.T, A.V.P., I.A. performed theoretical calculations, H.F., M.B., T.K., S.V., W.S. performed synthesis in the large volume press and the synthesis of precursors. B.M. and P.F. performed nanoindentation measurements. L.D. performed electrical resistance measurements. All authors contributed to the discussion of the results.

**Acknowledgements**

M.B. thanks the Deutsche Forschungsgemeinschaft (DFG project BY112/1-1). N.D. and L.D. thank the Deutsche Forschungsgemeinschaft (DFG projects DU 954-11/1 and DU 393-10/1) andthe Federal Ministry of Education and Research, Germany (BMBF, grant no. 5K16WC1) for financial support. S.V. and W.S. are grateful for a PhD fellowship granted by the Fonds der Chemischen Industrie (FCI), Germany. B.M. thanks the Deutsche Forschungsgemeinschaft (DFG project ME 4368/7-1) for financial support. Parts of this research were carried out at the Extreme Conditions Beamline (P02.2) at DESY, a member of Helmholtz Association (HGF). Portions of this work was performed at GeoSoilEnviroCARS (The University of Chicago, Sector 13), Advanced Photon Source (APS), Argonne National Laboratory. GeoSoilEnviroCARS is supported by the National Science Foundation - Earth Sciences (EAR - 1634415) and Department of Energy- GeoSciences (DE-FG02-94ER14466). This research used resources of the Advanced Photon Source, a U.S. Department of Energy (DOE) Office of Science User Facility operated for the DOE Office of Science by Argonne National Laboratory under Contract No. DE-AC02-06CH11357, as well as resources from the Center for Nanoanalysis and Electron Microscopy (CENEM) at Friedrich-Alexander University Erlangen-Nürnberg. Several high-pressure diffraction experiments were performed on beamline ID15B at the European Synchrotron Radiation Facility (ESRF), Grenoble, France. We thank Sven Linhardt and Gerald Bauer for the help with electrical resistance measurements.

# Supplementary Information

**High-pressure synthesis of ultraincompressible hard rhenium nitride pernitride Re$_2$(N$_2$)N$_2$ stable at ambient conditions**

Maxim Bykov, Stella Chariton, Hongzhan Fei, Timofey Fedotenko, Georgios Aprilis, Alena V. Ponomareva, Ferenc Tasnádi, Igor A. Abrikosov, Benoit Merle, Patrick Feldner, Sebastian Vogel, Wolfgang Schnick, Eran Greenberg, Vitali Prakapenka, Michael Hanfland, Anna Pakhomova, Hanns-Peter Liermann, Tomoo Katsura, Natalia Dubrovinskaia, Leonid Dubrovinsky

# Contents

## Analysis of multigrain/multiphase datasets

If a chemical reaction occurs in a DAC during laser heating, it often leads to multiple domains of well-crystallized phase(s). Moreover, the diffraction spots originating from different grains not often overlap with each other and this allows to integrate the dataset separately for each grain. The same happened in a case of the Re – $N_2$ system. For the unit cell determination, we use the reciprocal space viewer (Ewald explorer) from *CrysAlis$^{Pro}$*. It allows to manually select the reciprocal lattice of the grain. Below we provide an example for the typical dataset of $ReN_2$. The Supplementary Figure 1 demonstrates the reciprocal space reconstruction with seven strongest lattices corresponding to the grains of $ReN_2$ in different orientations. The Supplementary Table 1 shows typical statistics for total/overlapped reflections in our datasets.

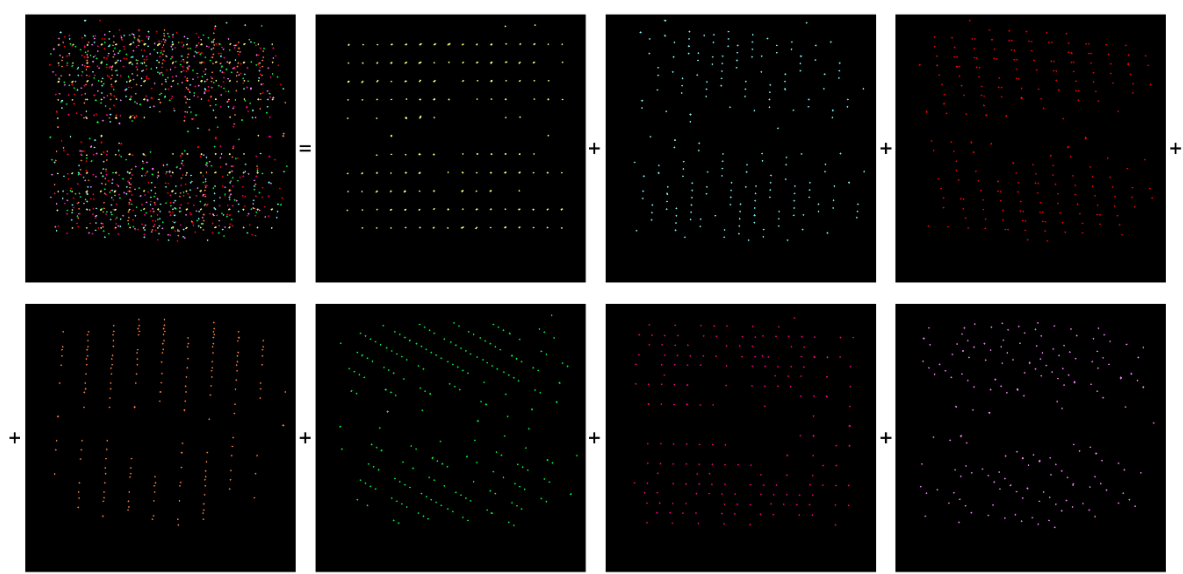

**Supplementary Figure 1**. Reciprocal space reconstruction with seven strongest lattices.

**Supplementary Table 1**. Typical reflection statistics for several $ReN_2$ grains, producing the strongest diffraction spots.

| Grain | $a$, Å | $b$, Å | $c$, Å | $\beta$, ° | $V$, Å$^3$ | Reflections total / separate / overlapped |
|---|---|---|---|---|---|---|
| #1 | 3.5065 | 6.1482 | 4.6884 | 111.175 | 94.3 | 249/249/0 |
| #2 | 3.5065 | 6.1544 | 4.6865 | 111.253 | 94.3 | 218/216/2 |
| #3 | 3.5072 | 6.1467 | 4.6921 | 111.197 | 94.3 | 249/246/3 |
| #4 | 3.5022 | 6.1615 | 4.6885 | 111.099 | 94.4 | 196/194/2 |
| #5 | 3.5020 | 6.1620 | 4.6906 | 111.214 | 94.4 | 243/237/6 |
| #6 | 3.5031 | 6.1528 | 4.6933 | 111.212 | 94.3 | 198/190/8 |
| #7 | 3.4967 | 6.1539 | 4.6939 | 111.149 | 94.2 | 208/199/9 |

## Estimation of the nitrogen content in the WC-type ReN$_x$

The ambient-pressure unit cell parameters and the unit cell volume of the WC-type ReN$_x$ phase, obtained in the experiments #1 and #7 do not agree with our and with previous theoretical calculations if $x = 1$ (Supplementary Table 2). Furthermore, the density of WC-ReN phase with ideal 1:1 stoichiometry from theoretical calculations is in a good agreement with the N-content – density trend of the Re-N system (Supplementary Figure 2a). However, the experimental density, calculated with the assumption of 1:1 composition, is clearly higher than expected (Supplementary Figure 2a). In order to estimate the nitrogen content, we have calculated the unit cell volumes of ReN$_x$ with $x$ varying from 0.58 to 1 (Supplementary Table 2, Supplementary Figure 2b). In the studied composition range, the dependence of the unit cell volume of ReN$_x$ linearly depends on the nitrogen content. The result of the linear fit, presented in the Supplementary Figure 2b was used to estimate the composition of ReN$_x$ obtained in the experiments #1 and #6.

We could not reliably detect ReN$_{0.6}$ phase in the experiment #2, which was used for the measurement of the equation of state. We would like to note here that the volume change of ReN$_{0.6}$ between 0 and 42 GPa from the available single-crystal data is only -9.75%, which makes it a very incompressible compound as well. Several theoretical works were devoted to the ReN compound and it was predicted that the most stable structure of ReN is NiAs like structure.[1,2] We believe that the role of vacancies in the stabilization of ReN$_x$ ($x \leq 1$) phases must be taken into account in further theoretical calculations.

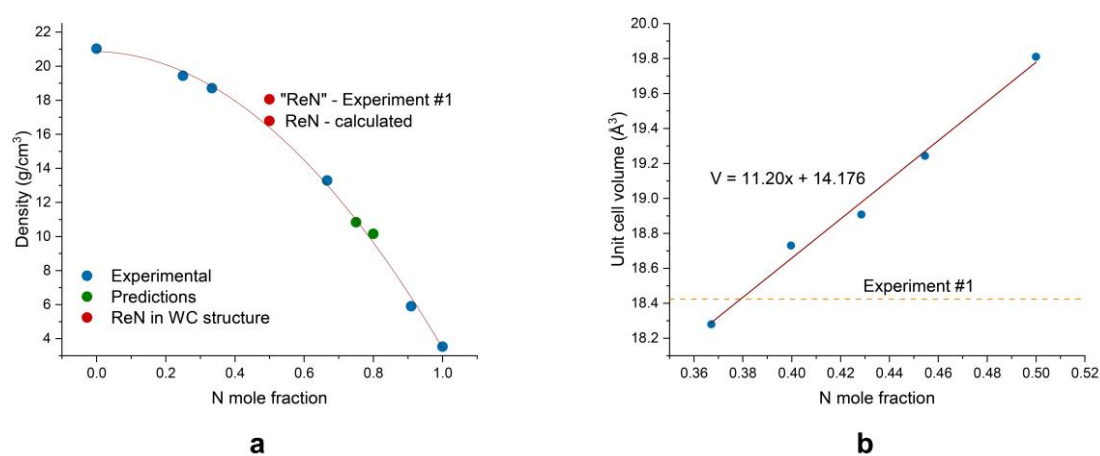

**Supplementary Figure 2.** (a) Density of Re-N compounds as a function of nitrogen content at ambient pressure. Blue points – experimental data on Re,[3] Re$_3$N,[4] Re$_2$N,[4] ReN$_2$ (this study), ReN$_{10}$,[5] $cg$-N.[6] Green points – theoretically predicted compounds ReN$_3$ and ReN$_4$.[7] Red points – calculated and experimental density of ReN in WC structure type. (b) Unit cell volume of WC-type Re-N phase as a function of nitrogen concentration. Blue points – calculated volumes of ReN$_{0.58}$, ReN$_{0.666}$, ReN$_{0.75}$, ReN$_{5/6}$ and ReN. Red line – linear fit.

**Supplementary Table 2.** Calculated and experimental lattice parameters for WC-type structure of ReN$_x$ at ambient pressure.

|  | $V$, Å$^3$ | Method | Reference |
|---|---|---|---|
| **Experiments:** | | | |
| Experiment #1 | 18.424 | SC XRD | This study |
| Experiment #6 | 18.19 | SC XRD | This study |
| **Calculations:** | | | |
| ReN | 19.46 | LDA TB-LMTO | [1] |
| ReN | 19.64 | GGA-PBE | [8] |
| ReN | 19.31 | LDA, FP-LAPW | [2] |
| ReN | 19.81 | PBEsol | This study |
| ReN$_{5/6}$ | 19.24 | PBEsol | This study |
| ReN$_{0.75}$ | 18.91 | PBEsol | This study |
| ReN$_{0.666}$ | 18.73 | PBEsol | This study |
| ReN$_{0.58}$ | 18.28 | PBEsol | This study |

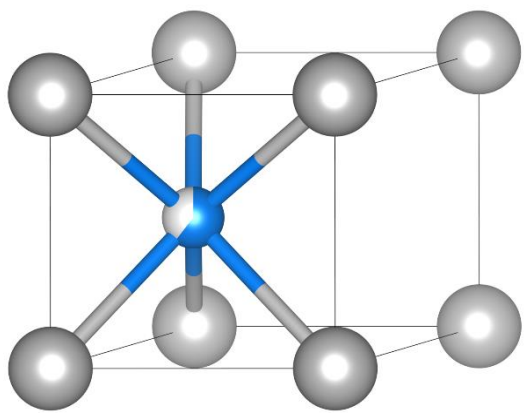

**Supplementary Figure 3**. The crystal structure of ReN$_{0.6}$ at ambient conditions. Re atoms – gray balls, N atoms – blue.

# Refinement and crystal structure details of ReN$_2$

**Supplementary Table 3.** Details on the refinement of the crystal structure of ReN$_2$ at ambient conditions

| Crystal data | |
|---|---|
| Chemical formula | ReN$_2$ |
| $M_r$, g/mol | 214.22 |
| Crystal system, space group | Monoclinic, $P2_1/c$ (No. 14) |
| Temperature (K) | 293 |
| Pressure (GPa) | 0.0001 |
| $a$, $b$, $c$ (Å) | 3.6254(17), 6.407(7), 4.948(3) |
| $\beta$ (°) | 111.48(6) |
| $V$ (Å$^3$) | 106.96(15) |
| $Z$ | 4 |
| Radiation type | Synchrotron, $\lambda = 0.2903$ Å |
| $\mu$ (mm$^{-1}$) | 10.76 |
| Crystal size (mm$^3$) | $0.03 \times 0.02 \times 0.01$ |
| **Data collection** | |
| Diffractometer | 13IDD, APS, Chicago, USA |
| Absorption correction | Multi-scan |
| $T_{min}$, $T_{max}$ | 0.386, 1.000 |
| No. of measured, independent and observed [$I > 2\sigma(I)$] reflections | 290, 180, 175 |
| $R_{int}$ | 0.011 |
| $(\sin\theta/\lambda)_{max}$ (Å$^{-1}$) | 0.879 |
| **Refinement** | |
| $R[F^2 > 2\sigma(F^2)]$, $wR(F^2)$, $S$ | 0.035, 0.088, 1.10 |
| No. of reflections | 180 |
| No. of parameters | 19 |
| $\Delta\rho_{max}$, $\Delta\rho_{min}$ (e Å$^{-3}$) | 2.83, -2.67 |
| **Refined crystal structure** | |
| Re ($x$, $y$, $z$) | 0.35490(11), 0.34041(8), 0.19965(8) |
| N1 ($x$, $y$, $z$) | 0.194(2), 0.038(2), 0.311(19) |
| N2 ($x$, $y$, $z$) | 0.259(3), 0.6381(18), 0.024(2) |
| **Calculated crystal structure** | |
| Re ($x$, $y$, $z$) | (0.35397, 0.33961, 0.19931) |
| N1 ($x$, $y$, $z$) | (0.1889, 0.037, 0.30) |
| N2 ($x$, $y$, $z$) | (0.2540, 0.6397, 0.0164) |

**Supplementary Table 4**. Selected geometric parameters of ReN$_2$ at ambient conditions (Å)

| | | | |
|---|---|---|---|
| Re01—Re01$^i$ | 2.7318 (16) | Re01—N2$^{vi}$ | 2.095 (10) |
| Re01—Re01$^{ii}$ | 2.7318 (16) | N1—Re01$^{iv}$ | 2.112 (11) |
| Re01—N1 | 2.105 (12) | N1—Re01$^{ii}$ | 2.082 (9) |
| Re01—N1$^{iii}$ | 2.112 (11) | N1—N1$^{vii}$ | 1.412 (16) |
| Re01—N1$^i$ | 2.082 (9) | N2—Re01$^v$ | 2.082 (10) |
| Re01—N2 | 2.072 (11) | N2—Re01$^{vi}$ | 2.095 (9) |
| Re01—N2$^{iv}$ | 2.028 (11) | N2—Re01$^{iii}$ | 2.028 (11) |
| Re01—N2$^v$ | 2.082 (10) | | |

Symmetry code(s): (i) $x$, $-y+1/2$, $z+1/2$; (ii) $x$, $-y+1/2$, $z-1/2$; (iii) $-x+1$, $y+1/2$, $-z+1/2$;

(iv) $-x+1$, $y-1/2$, $-z+1/2$; (v) $-x+1$, $-y+1$, $-z$; (vi) $-x$, $-y+1$, $-z$; (vii) $-x$, $-y$, $-z$.

## Compressibility of ReN$_2$

**Supplementary Table 5**. Experimental lattice parameters of ReN$_2$ on compression.

| Pressure, GPa | $a$, Å | $b$, Å | $c$, Å | $\beta$, ° | $V$, Å$^3$ |
|---|---|---|---|---|---|
| 0.0001 | 3.62565(16) | 6.4216(2) | 4.9478(3) | 111.446(4) | 107.218(9) |
| 3.37(5) | 3.6171(2) | 6.4118(4) | 4.9234(3) | 111.250(5) | 106.420(12) |
| 4.10(5) | 3.6154(3) | 6.4106(6) | 4.9161(6) | 111.279(8) | 106.173(19) |
| 6.95(9) | 3.6092(4) | 6.3961(7) | 4.9050(6) | 111.256(8) | 105.53(2) |
| 8.97(6) | 3.6055(3) | 6.3872(8) | 4.8974(6) | 111.267(8) | 105.10(2) |
| 11.18(5) | 3.5993(4) | 6.3770(8) | 4.8822(6) | 111.261(11) | 104.43(2) |
| 12.79(11) | 3.5957(4) | 6.3723(8) | 4.8715(9) | 111.242(12) | 104.03(3) |
| 14.74(11) | 3.5900(4) | 6.3623(8) | 4.8617(8) | 111.177(12) | 103.54(3) |
| 17.97(13) | 3.5849(5) | 6.3517(8) | 4.8523(6) | 111.202(11) | 103.01(2) |
| 25.67(15) | 3.5728(6) | 6.3071(10) | 4.8226(8) | 111.243(14) | 101.29(3) |
| 32.11(17) | 3.5604(6) | 6.2742(10) | 4.7992(8) | 111.264(14) | 99.91(3) |
| 39.37(13) | 3.5482(9) | 6.2338(13) | 4.7735(9) | 111.272(17) | 98.39(4) |
| 45.44(14) | 3.5379(9) | 6.1963(10) | 4.7503(8) | 111.368(15) | 96.98(4) |

## Representative powder diffraction patterns. Synthesis of ReN$_2$ *via* a reaction with nitrogen

*Note 1:* Due to the small lattice parameters of defect WC-type ReN$_x$, we could not reliably detect it on some powder patterns, where it could be in a mixture with ReN$_2$ and Re$_2$N. Single-crystalline grains of ReN$_x$ were found only in the experiments #1 and #6, but we cannot exclude that this phase may be present in other syntheses.

*Note 2:* Some powder diffraction patterns contain very weak non-indexed peaks. These peaks may originate from the pressure-transmitting medium (N$_2$) or other minor phases. Le Bail fits were performed on the major phases, which have also been confirmed by the single-crystal experiments. Peak indexing of weak unknown phases based on the powder pattern in such a mixture would be unreliable.

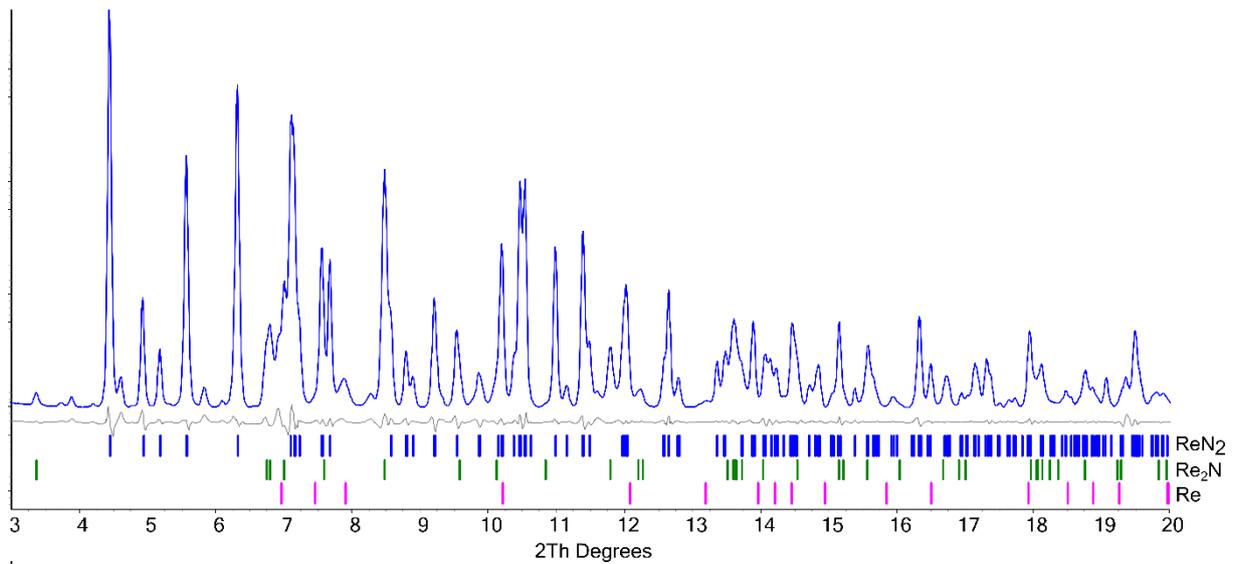

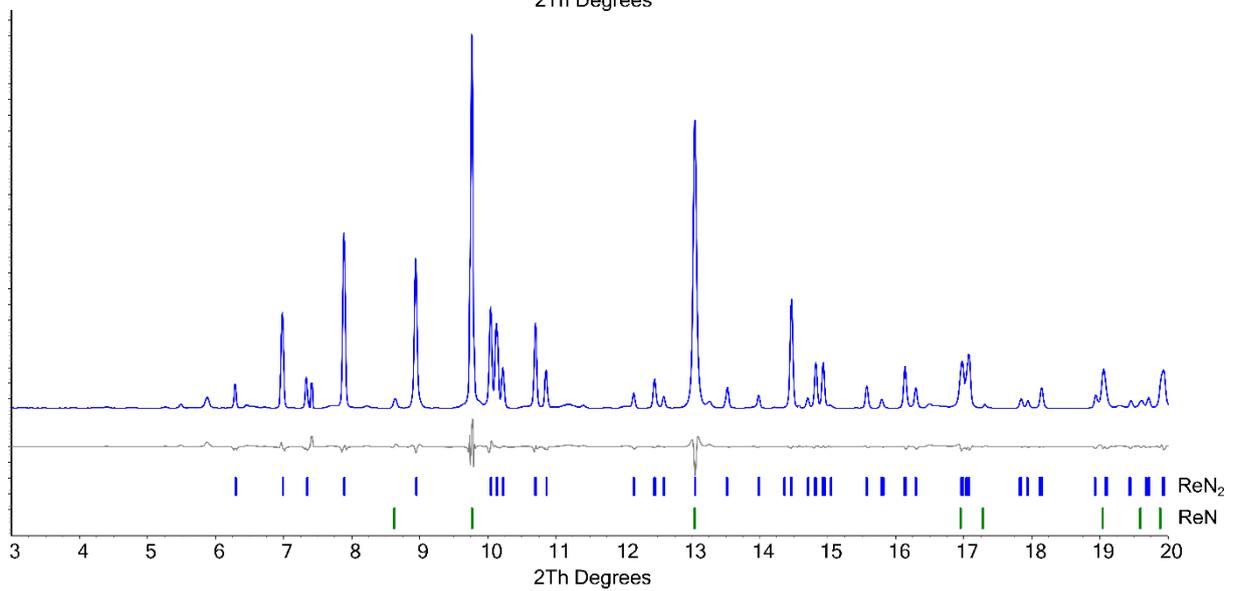

**Supplementary Figure 4**. Powder diffraction pattern of the sample #1 at ambient pressure ($\lambda$ = 0.29 and 0.41 Å for upper and lower patterns respectively).

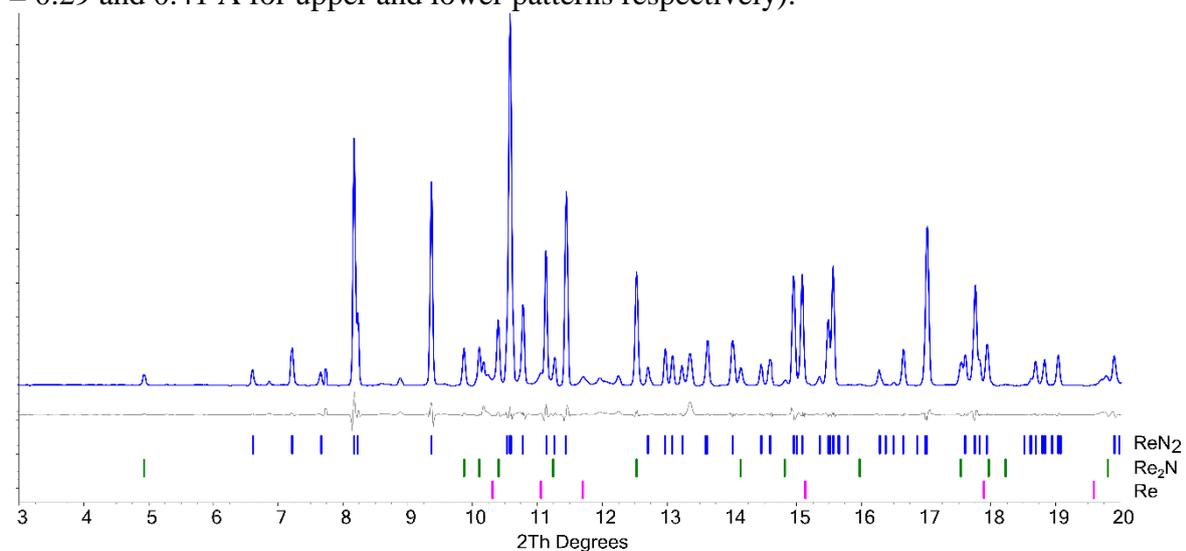

**Supplementary Figure 5**. Powder diffraction pattern of the sample #3 at 71 GPa ($\lambda$ = 0.41 Å).

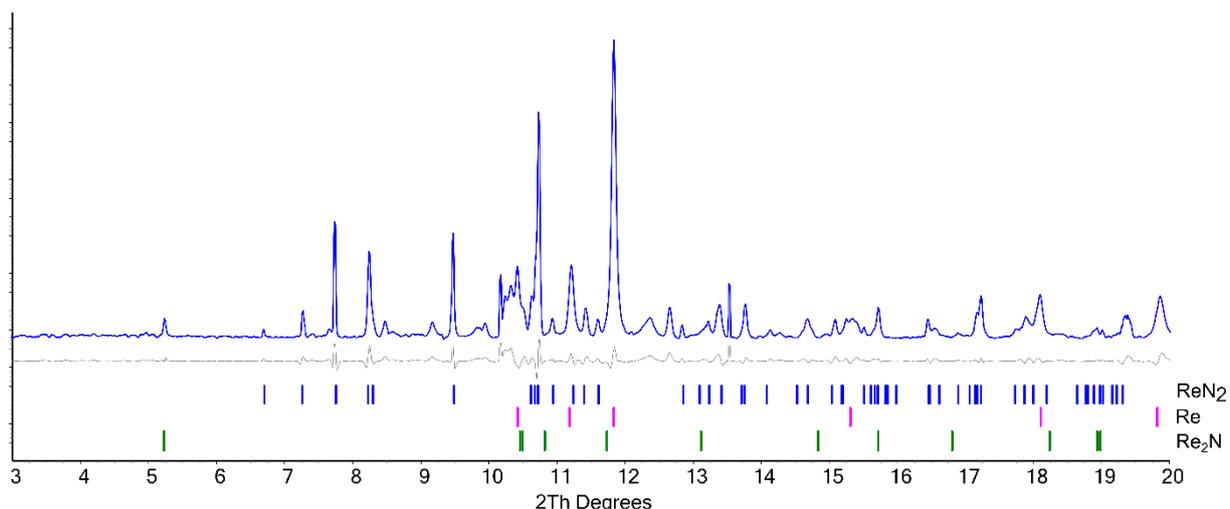

**Supplementary Figure 6**. Powder diffraction pattern of the sample #4 at 86 GPa (λ = 0.41 Å).

## Choice of solid nitrogen precursors

The reaction between Re and NaN$_3$ resulted in the main product, which has hexagonal symmetry P6$_3$/*mmc* (*a* = 2.7715, *c* = 11.2191 Å). The structure solution and refinement revealed its chemical formula as NaReN$_2$. The structure is based on layered MoS$_2$-type structure with sodium atoms intercalated between ReN$_2$ layers (Supplementary Figure 7). Na occupies the crystallographic site 2*a* (0, 0, 0), Re – 2*c* (1/3, 2/3, 1/4), N – 4*f* (1/3, 2/3, 0.6405). It should be noted that the lattice parameters of this phase are very close to those of the ReN$_2$ phase, that was synthesized by Kawamura *et al.*,[9] with a difference that the lattice parameter *c* of NaReN$_2$ is larger. If we consider the reaction, reported by Kawamura *et al.*:[9]

$$\text{ReCl}_5 + 2\text{Li}_3\text{N} + x\text{NaCl} \xrightarrow{7.7\ GPa,\ 1200°C} \text{ReN}_2 + 5\text{LiCl} + x\text{NaCl} \quad (1),$$

we can notice two important problems: The equation is not balanced (and cannot be balanced, even if we consider the release of free nitrogen on the right side of the equation) and rhenium is the only element that has changed its oxidation state. We suggest that the real product of the reaction (1) is LiReN$_2$. This explains both issues mentioned above, and is in agreement with shorter *c*-axis than in NaReN$_2$ due to the smaller cation radius of lithium compared to sodium.

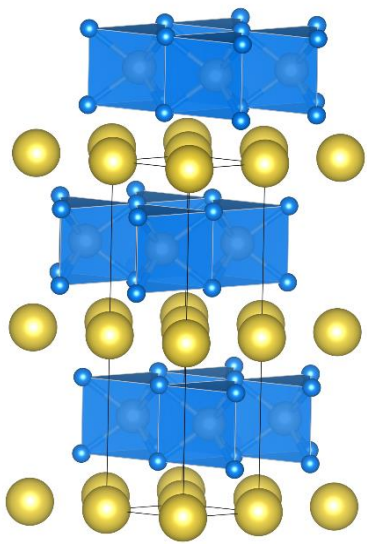

**Supplementary Figure 7**. The crystal structure of NaReN$_2$ from the synthesis #5 at ~40 GPa. Yellow balls – Na atoms. Blue polyhedra – ReN$_6$ trigonal prisms. Re occupies the site 2c (1/3, 2/3, ¼), N – 4f (1/3, 2/3, 0.64217), Na – 2a (0 0 0).

Sodium azide appeared to be not a suitable source of nitrogen for high-pressure nitridation reaction, but opens a route to ternary nitrides with intercalated alkali metals, which in turn may be important for the development of high-performance electrode materials.[10]

The experiment in LHDAC with NH$_4$N$_3$ as a source of nitrogen (Experiment #6, Table 1) resulted in the synthesis of ReN$_2$ and defect WC-type ReN$_x$. The unit cell volume of WC-type ReN$_x$ at ambient pressure appeared to be slightly smaller than that in the Experiment#1 [$V_{exp1}$ = 18.42(1) Å, $V_{exp6}$ = 18.19(1) Å$^3$] and the estimated composition of this compound is ReN$_{0.56}$. The Experiments #6 and 7 show that NH$_4$N$_3$ precursor may be successfully used for the synthesis of rhenium nitrides when a solid source of nitrogen is required.

Currently, several methods are used for the synthesis of nitrides in the LVP. One route is a high-pressure solid-state metathesis (HPSSM) reaction between an oxidized metal precursor and a nitride (*e.g.* boron nitride BN or lithium nitride Li$_3$N).[9,11–14] Recently Lei *et al*. reported a novel synthetic route to rhenium nitride Re$_3$N:[13,15]

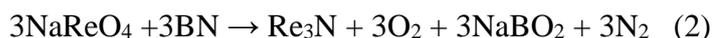
$$3NaReO_4 + 3BN \rightarrow Re_3N + 3O_2 + 3NaBO_2 + 3N_2 \quad (2)$$

Although, the HPSSM method resulted in a number of exciting discoveries, it has a few disadvantages. In case of a simple metathesis reaction, where the oxidation state of elements is not changed it will not be possible to obtain compounds with N-N units. In case of redox reactions, where metal itself serves as an oxidizer of nitrogen, it is hard to create an excess of nitrogen that can prevent decomposition of target nitrogen-rich phases. Furthermore, there are always several side products of such a reaction: *e.g.* the release of free oxygen like in the reaction (3) may influence the reaction in some cases. Schnick *et al.* successfully used controlled decomposition of azides to obtain diazenides $BaN_2$, $SrN_2$ and $CaN_2$ as well as $Li_2N_2$ in a large volume press.[16,17] This is a much cleaner method than a metathesis reaction, but is demanding, because not all metal azides are readily available and safe to work with. $NH_4N_3$ appears to be a good choice for the synthesis of binary nitrides of transition metals due to several reasons: It has a high content of nitrogen (93.3 wt. %), and it can serve as an oxidizer, so that the elemental metal can be used for the reaction. Compared with the metal azides, it is relatively safe to work with this material. The dissociation of the excess of $NH_4N_3$ may create high partial pressure of $N_2$, which prevents the decomposition of target nitrogen-rich phases (Le Chatelier principle).[18]

## Synthesis of ReN₂ in a multianvil apparatus

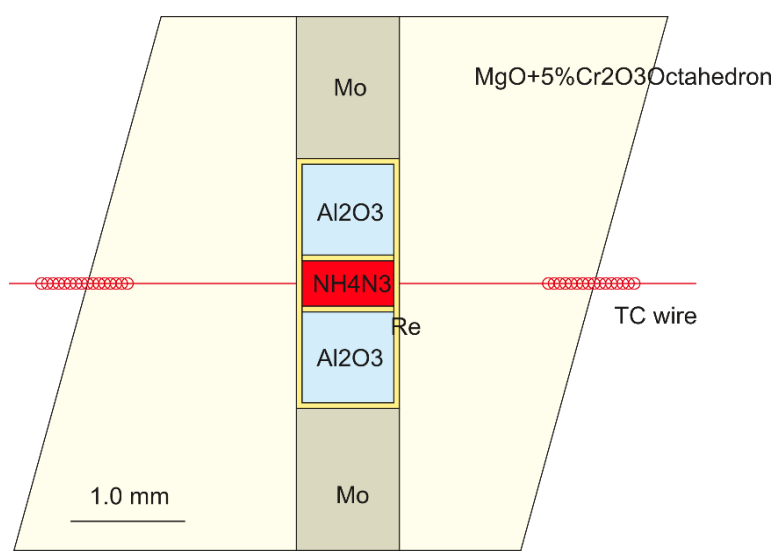

**Supplementary Figure 8.** Schematic drawing of high-pressure cell assembly for the multianvil synthesis of rhenium nitrides.

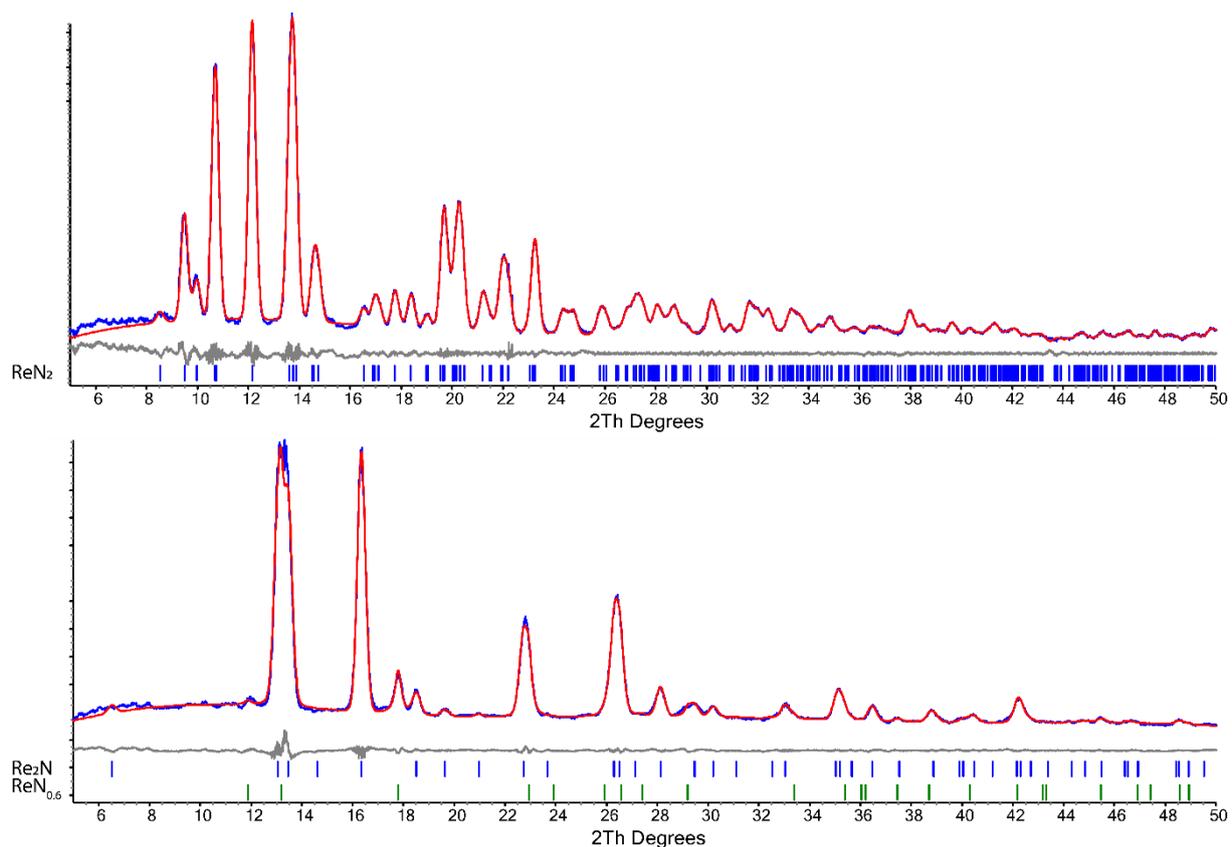

**Supplementary Figure 9.** Powder diffraction pattern of the samples recovered from the synthesis in the multianvil press (Experiment #7), λ = 0.56 Å (Ag-Kα). $ReN_2$ and $Re_2N$ phases were characterized by single-crystal X-ray diffraction, while $ReN_{0.6}$ is evidenced only from the powder XRD.

## Supplementary references